\documentclass[12pt,a4paper]{article}

\usepackage[utf8]{inputenc}
\usepackage[T1]{fontenc}
\usepackage{lmodern}

\usepackage[english]{babel}
\usepackage{csquotes}

\usepackage{amsmath}
\usepackage{amssymb}

\usepackage[margin=1in]{geometry}
\usepackage{setspace}
\usepackage[hidelinks]{hyperref}

\usepackage{titlesec}

\usepackage{abstract}

\usepackage{xcolor}
\usepackage{tikz}
\usetikzlibrary{arrows.meta,positioning,shapes.geometric,shapes.misc,calc}

\title{\textbf{The Ethical Knowledge Gap}\\[0.5em]
\large Dispersed Knowledge, Sensemaking Failures, and Epistemic Dependence}

\author{Jan Gogoll}
\date{}

\begin{document}

\maketitle

\begin{abstract}
\noindent
Ethical software development remains stubbornly difficult despite two decades of normative frameworks, professional codes, and participatory methodologies. This paper offers a diagnostic rather than prescriptive contribution: it argues that the persistent gap between ethical intention and ethical implementation is a structural epistemic condition, not primarily a failure of will, education, or normative guidance. Three independently sufficient mechanisms interact to produce what I call the \emph{ethical knowledge gap}---a condition in which the knowledge required for ethically informed decision-making is systematically unavailable at the point of decision, even when the organization as a whole possesses it. First, drawing on Hayek's (1945) analysis of dispersed knowledge and its organizational extensions, the paper establishes that ethically relevant knowledge in software development is constitutively distributed across roles, largely tacit, and---unlike efficiency-related knowledge---unsupported by any spontaneous aggregation mechanism analogous to the price system. Second, an \emph{interpretive deficit}, analyzed through Weick's sensemaking framework and the literature on framing and epistemic cultures, renders developers unable to recognize the ethical significance of what they know: the sensemaking apparatus of engineering culture makes technical decisions intelligible while systematically obscuring their ethical dimensions. Third, a \emph{credibility attenuation}, analyzed through the social epistemology of testimony and epistemic dependence, discounts developers' observations as they cross organizational role boundaries, so that hybrid judgments combining technical detail with ethical assessment lose their epistemic force. A formal sketch models the cumulative effect of these mechanisms and supports a structural blockage claim: no single-vector intervention---ethics training, review boards, reporting channels, or participatory formats alone---can close the gap, because each addresses at most one mechanism while leaving the others intact. The paper derives design principles for a \emph{translation layer} that must simultaneously enrich sensemaking categories, restructure epistemic-dependence architectures, and enable targeted knowledge aggregation. The framework also reveals that the gap is symmetric: the same structural conditions that prevent organizations from recognizing ethical harms equally prevent them from recognizing ethical opportunities.\end{abstract}

%---------------------------------------------------------------------
\section{Introduction}
%---------------------------------------------------------------------

The ethics of software development is by no means a neglected topic. Over the past two decades, a substantial body of work has developed normative frameworks, methodologies, and institutional proposals for making software production more ethically responsive. Value-sensitive design, for instance, offers a tripartite methodology for integrating human values into system design from the outset (Friedman and Hendry 2019). Responsible innovation provides a governance framework organized around anticipation, reflexivity, inclusion, and responsiveness (Stilgoe, Owen, and Macnaghten 2013). Process-oriented approaches have proposed embedding ethical deliberation directly into development workflows, for instance through structured identification, deliberation, and embedding phases tied to agile ceremonies and the Product Owner role (Gogoll and Zuber 2026). Professional codes of ethics articulate principles---fairness, transparency, accountability, privacy---that are by now widely recognized, if unevenly implemented. Taken together, this literature provides significant normative resources: there is no shortage of principled guidance on which values matter, how they might be identified, and when they should be considered.

And yet ethical software development remains stubbornly difficult in practice. The gap between principled commitment and actual implementation is not merely anecdotal but is empirically documented across multiple studies and organizational settings. Madaio et al.\ (2020) found that fairness work in AI organizations depends on ad hoc advocacy by motivated individuals and becomes durable only when supported by organizational infrastructures that most organizations lack. Holstein et al.\ (2019) showed that industry practitioners working on ML fairness encounter barriers that are not primarily technical but organizational: unclear responsibility allocation, limited decision rights, and misalignment between ethical concerns and existing workflows. Rakova et al.\ (2021) documented a persistent disconnect between ethical principles endorsed at the organizational level and their translation into concrete design and implementation decisions. Widder and Nafus (2023) found that modular system architecture encourages engineers to displace ethical responsibility along the AI supply chain---upstream, downstream, or to another team---so that harms are recognized in principle but systematically unaddressed in practice. The pattern across these findings is striking: practitioners frequently report awareness that ethical issues exist while lacking actionable pathways to connect that awareness to their everyday technical work. Organizations adopt ethical frameworks, articulate values, and express genuine commitment---and the ethical concerns still fail to shape the decisions that determine what the software actually does.

The persistence of this gap is not for lack of trying. Organizations have responded with a range of interventions that, while valuable in their own right, consistently fall short of closing it. Ethics courses in computer science curricula may expand individual moral vocabulary but do not restructure the organizational categories through which technical decisions are classified once graduates enter the workplace. Cross-functional workshops and alignment meetings bring relevant actors into the same room, but the high-level principles they produce often remain too abstract to connect to the implementation-level decisions where ethical issues actually reside. Corporate value statements articulate commitments that are sincere and publicly visible, yet add another layer of principle to a landscape already rich in principles while leaving untouched the structural conditions that prevent principles from reaching practice. Each of these interventions addresses something real: a deficit in individual knowledge, a lack of cross-functional contact, an absence of explicit normative commitment. But the empirical pattern suggests that the problem lies deeper than any of them reaches on its own. The knowledge exists (somewhere), the goodwill exists, the principles exist, yet, the gap persists.

This paper asks why. Its contribution is not another normative framework for ethical software development but a diagnostic one. It argues that the persistent gap between ethical intention and ethical implementation is not primarily a failure of will, education, or normative guidance, but a structural epistemic condition rooted in how (software) organizations produce and coordinate knowledge. Specifically, three mechanisms interact to produce what may be called the \emph{ethical knowledge gap}: a structural condition in which the knowledge required for ethically informed decision-making is systematically unavailable at the point of decision, even when the organization as a whole may well possess it.

The first mechanism, analyzed through Hayek's (1945) account of dispersed knowledge and its organizational extensions (Tsoukas 1996; Grant 1996; Foss 1999), is the constitutive dispersal of (ethically) relevant knowledge across organizational roles. Developers hold implementation knowledge---how the system actually behaves at a granular level, including edge cases, silent failures, data persistence patterns, and design trade-offs. Managers, product owners, and domain experts hold contextual knowledge---who will use the system, under what conditions, with what vulnerabilities, and in what regulatory and social environment. The ethical assessment of any given development decision typically requires both kinds of knowledge simultaneously. Yet no single agent usually holds both, and much of the relevant knowledge (particularly on the implementation side) is tacit, local, and resistant to propositional formulation.

The second mechanism is a structural \emph{interpretive deficit}. Even when ethically relevant knowledge is accessible to an agent, it may fail to be recognized as ethically significant. The sensemaking apparatus of engineering culture, which includes its categories, vocabularies, and routines, renders technical work intelligible while systematically obscuring its ethical dimensions. The developer who knows that a caching layer retains data for 72 hours after deletion encounters this primarily as a technical design trade-off, not as an ethical decision about user privacy, because the interpretive categories available to them do not include the pathway from implementation detail to ethical significance.

The third mechanism is \emph{credibility attenuation}. Even when a developer does perceive and articulate an ethical concern, the organization's epistemic-dependence architecture, implemented to assign expertise along professional-role lines, may discount the testimony as it crosses domain boundaries. The developer is credited as a speaker on implementation but not necessarily on the ethical implications of that implementation. The integrated observation, which combined technical knowledge with contextual judgment, gets disaggregated, reframed, deferred, or dismissed.

These three mechanisms are independently recognized in the literature, but their interaction in the context of software development has not been analyzed. The central claim of this paper is that the ethical knowledge gap is sustained by all three simultaneously, and that this explains why single-vector interventions---ethics training alone, review boards alone, reporting channels alone, encouragement to ``speak up'' alone---consistently underperform. Each addresses at most one mechanism while leaving the others intact. What is needed is an institutional design that counteracts dispersal, the interpretive deficit, and credibility attenuation at the same time.

%---------------------------------------------------------------------
\section{Existing Frameworks and Their Limitations}
%---------------------------------------------------------------------

\subsection{Principal--Agent Theory}

The most established formal treatment of information asymmetry in economics is principal--agent theory (Akerlof 1970; Spence 1973; Stiglitz 1975). In its standard form, a principal delegates a task to an agent whose actions, effort, or private information cannot be fully monitored. The resulting problem is one of incentive alignment under asymmetric information: because the agent knows something the principal does not, or can act in ways the principal cannot directly observe, the principal must design contracts, monitoring arrangements, or signaling mechanisms to induce behavior that serves the principal's interests---the common example being shareholders delegating to management.

This framework is illuminating where the central challenge is opportunism, moral hazard, or adverse selection. Yet it is poorly suited to the ethical knowledge problem in software development. First, the asymmetry at issue here is not unidirectional. In standard principal--agent settings, the principal is modeled as the locus of authority and objective-setting, while the agent holds private information relevant to implementation. In ethical software development, by contrast, relevant knowledge is dispersed across multiple roles and organizational levels: developers possess technical and situational knowledge, managers possess strategic and organizational knowledge, domain experts possess contextual knowledge, users and affected stakeholders possess experiential knowledge, and no single actor fully sees the ethically relevant picture. The problem is therefore not that one side knows what the other side hides, but that the knowledge required for responsible judgment is fragmented across the organization.

Second, principal--agent theory is primarily concerned with motivational divergence. It assumes that actors may fail to disclose or act on information because their incentives are misaligned. In the ethical case, however, the difficulty often persists even in the absence of bad faith or strategic withholding. A developer may be willing to raise concerns, a manager may be willing to listen, and yet ethically relevant knowledge still fails to enter decision-making because it remains local, partial, weakly articulated, or institutionally unintegrated. The central obstacle is therefore not merely incentive misalignment, but the organizational dispersion and practical inaccessibility of knowledge.

Third, principal--agent theory usually treats information as something that can, at least in principle, be transferred if the right contractual or monitoring mechanism is found. By contrast, a Hayekian perspective emphasizes that much of the knowledge relevant to software ethics is tacit, situated, and dependent on ``particular circumstances of time and place'' (Hayek, 1945). Such knowledge is often embedded in professional experience, domain familiarity, user contact, or local workflow conditions, and cannot simply be exhaustively stated in propositional form and passed upward through the hierarchy. The issue is thus not only that information is asymmetrically distributed, but that some of it resists formalization altogether.

For these reasons, the ethical knowledge problem in software organizations is better understood not as a classic principal--agent problem, but as a problem of distributed, partly tacit, and organizationally fragmented knowledge. What is needed is not merely better incentives for disclosure, but structures that enable the identification, articulation, integration, and uptake of ethically relevant knowledge across roles.

\subsection{Knowledge-Based Theories of the Firm}

A more promising starting point lies in the organizational knowledge literature that has, since the 1990s, explicitly imported Hayekian insights into theories of the firm. Tsoukas (1996) reconceived the firm as a ``distributed knowledge system'' in which knowledge is socially embedded, partly tacit, and cannot simply be documented or centralized. Grant (1996) developed a knowledge-based theory of the firm in which the central organizational challenge is the integration of specialist knowledge held by individuals. This knowledge remains primarily located in those individuals even as it is coordinated through organizational mechanisms. Foss (1999) argued that modern organizational economics systematically underestimates coordination problems by failing to account for the genuine dispersal of knowledge that Hayek identified. Jensen and Meckling (1995) connected dispersed ``specific knowledge'' to the allocation of decision rights: because specific knowledge is costly to transfer, decision authority should be located where the knowledge sits, though this in turn generates control problems. Simon (1991) reinforced this picture by arguing that modern economies are fundamentally ``organizational economies'' and that organizational theory is as necessary as market theory for understanding coordination.

These contributions establish that the knowledge problem inside organizations is a recognized and well-studied phenomenon. The knowledge-based theory of the firm literature treats knowledge dispersal as a coordination and efficiency problem. This article's contribution is to show that when the dispersed knowledge is ethically relevant, two additional mechanisms---one interpretive, one credibility-related---compound the coordination failure in ways that the existing literature has not analyzed.

\subsection{Other Frameworks}

The Collingridge dilemma (1980) captures an important temporal dimension regarding technology production: a technology's consequences cannot be predicted with certainty until it is widely deployed, but by then it is too entrenched to change as it becomes too costly both monetarily and in terms of user acceptance once users have grown accustomed to it and created social practices around this technology. This identifies a timing problem without explaining the epistemic mechanisms that produce it.

Thompson's (1980) ``problem of many hands'' formalizes how moral responsibility diffuses in collective decision-making. The present paper can be read as providing an epistemic foundation for Thompson's observation: the reason no one acts is not merely that responsibility is diffuse but that the knowledge required for morally motivated action is structurally unavailable to any single agent.

Value-sensitive design (Friedman and Hendry 2019) is probably the most widespread and developed methodological framework for integrating ethical values into technology design. Its tripartite methodology---conceptual, empirical, and technical investigations---provides a structured process for identifying stakeholders, eliciting their values, and translating those values into design requirements. VSD's emphasis on participatory engagement and its explicit attention to indirect stakeholders represent genuine advances over purely technical design processes. Yet, VSD operates with a characteristic directionality that limits its reach with respect to the problem analyzed in this paper. The methodology proceeds top-down: it begins with the identification of relevant values---privacy, fairness, autonomy, human welfare---and then works downward toward design specifications through what the literature calls ``values hierarchies'' (van de Poel 2013). The ethical knowledge gap, by contrast, is a bottom-up problem. Ethically relevant facts arise at the level of implementation---a caching decision, a training data filter, a silent failure mode---and must be recognized as such before any value framework can be applied to them.

This difference in directionality has a structural consequence. VSD can identify that ``fairness'' or ``freedom from bias'' are relevant values for an insurance pricing algorithm. What it cannot do, by its own methodology, is surface the fact that a geocoding step in the data pipeline encodes census-block-level demographic composition into risk scores---because that fact is an implementation detail held locally by a data engineer who classifies it as a routine feature enrichment problem, not a fairness problem. The participatory formats that VSD and others recommend (e.g.\ stakeholder workshops, co-design sessions, empirical investigations) address the logistics of knowledge aggregation by bringing diverse actors into the same deliberative space. But assembling the relevant actors does not by itself ensure that implementation-level knowledge becomes ethically legible or ethically salient to those who hold it, nor that it is taken seriously when communicated across professional role boundaries. Participatory formats create the opportunity for ethically relevant knowledge to surface, but they do not address the prior question of whether participants can recognize the ethical significance of what they know, or whether the organizational setting treats their observations as credible contributions to ethical judgment rather than as technical input outside their authorized domain. As Sections~4 and~5 will argue in detail, both of these---the recognizability of ethical significance (4.1) and the credibility of cross-domain testimony (4.2)---face structural obstacles that participatory formats alone do not dissolve. The translation gap between abstract values and concrete design requirements has been widely noted as a persistent weakness of VSD in practice (Manders-Huits 2011; Wong et al.\ 2023). The argument developed here suggests that this gap is not merely a methodological shortcoming to be addressed by better tools or more careful facilitation. It is a consequence of the fact that VSD presupposes the very epistemic conditions---ethical recognizability and credible cross-boundary communication---that the ethical knowledge gap structurally undermines.

None of this implies that participatory approaches are misguided or dispensable. On the contrary, any adequate response to the ethical knowledge gap will necessarily include participatory elements, since the dispersal of knowledge across roles makes cross-boundary exchange indispensable. The claim is rather that participation is necessary but not sufficient. What is additionally required---and what VSD does not provide---is an account of the structural conditions under which participatory exchange actually succeeds in surfacing and integrating ethically relevant knowledge. The present paper aims to supply that account and, in doing so, to identify the epistemic preconditions that participatory approaches such as VSD implicitly assume to be in place.

%---------------------------------------------------------------------
\section{The Hayekian Layer: Knowledge as Constitutively Dispersed}
%---------------------------------------------------------------------

\subsection{Hayek on Knowledge}

F.\,A.\ Hayek's account of knowledge emerged in the context of the socialist calculation debate, that is, the dispute over whether a centrally planned economy could rationally allocate resources without (real) market prices. In ``The Use of Knowledge in Society'' (1945), Hayek argued against the feasibility of central economic planning on epistemic grounds. His point was not merely that centralized planning might be inefficient in practice, but that it is constrained by a deeper epistemic problem: the knowledge required for rational allocation is never available in its totality to any single mind. Rather, it is dispersed across society in incomplete, often contradictory fragments. The spirit of the argument may best be captured in a statement Hayek made elsewhere: ``The curious task of economics is to demonstrate to men how little they really know about what they imagine they can design'' (Hayek, 1988).

In the following I offer only a condensed reconstruction of Hayek's argument. Its full complexity cannot be developed within the space constraints of this article, and readers are encouraged to turn to the original essay for a more complete understanding. His argument rests on a set of interrelated conditions: (i)~the data required for rational allocation are never given to a single mind, but distributed across many actors; (ii)~a large portion of this knowledge consists of \emph{knowledge of particular circumstances of time and place}. It is local, situational, embedded in practice, often tacit, and resistant to formulation as general rules; (iii)~the relevant environment is dynamic, so that centralization faces not merely problems of collection, but also problems of reaction time; (iv)~the price system functions as a unique coordination mechanism that allows individuals with little aggregate knowledge to make compatible decisions because prices compress information about scarcity and demand and do so efficiently; and (v)~this mechanism presupposes an institutional environment, including competition and property rights, that makes decentralized planning feasible.

Hayek's argument thus combines an epistemic diagnosis with an institutional claim. The epistemic diagnosis is that socially relevant knowledge is dispersed, situated, and often changes too quickly to be centralized and used effectively. The institutional claim is that the price system enables coordination without requiring any actor to possess an overview of the whole (assuming defined property laws and individual freedom). Individuals need not know the total state of the economy, they need only respond to signals that register changes in relative scarcity and demand. Decentralization is therefore, on Hayek's view, not simply one organizational option among others, but an epistemic necessity imposed by the nature of knowledge itself.

\subsection{Organizational Transfer}

Since even highly complex organizations are necessarily subsets of the broader economy, the conditions of knowledge within them cannot simply be assumed to mirror those of society at large. We therefore need to recast the argument from ``knowledge in society'' as one about ``knowledge in organizations.'' The transfer of these Hayekian conditions to organizations has been extensively examined (Foss 1999; Foss and Klein 2013). Conditions (i) through (iii) apply with considerable force to knowledge-intensive organizations, and to software development in particular. Decision-relevant knowledge is distributed across specialist roles and organizational sites with much of it remaining local to concrete practices, systems, and workflows while the underlying environment changes continuously. In software development, dependencies between components are volatile and often emerge locally, while knowledge about how a system actually behaves is scattered across development, operations, security, support, UX, and other functions. At the same time, user behavior, production environments, regulatory expectations, and economic/competition landscapes are dynamic, so that central coordination faces not only information deficits but also timing problems (Cataldo et al.\ 2006; Herbsleb and Mockus 2003). Foss and Foss (2006) therefore argue that even within organizations, Hayekian knowledge dispersal produces genuine coordination problems that cannot be eliminated by hierarchy alone.

However, conditions (iv) and (v) do not transfer straightforwardly and require modification in the organizational context. Inside organizations, there are no genuine market prices. At best, they are approximated by budgets, KPIs, internal transfer pricing, or quasi-market structures (Coase 1937; Williamson 1991). Yet these devices do not replicate the spontaneous and anonymous information-aggregating function of prices in markets. The organization therefore lacks a fully equivalent mechanism for coordinating dispersed knowledge---as there is no such thing as a straightforward currency for ethical deliberation. The absence of a price-like mechanism is especially consequential for ethically relevant knowledge. For coordination and efficiency problems, organizations possess imperfect but relatively functional substitutes: budgets register resource scarcity, KPIs compress performance information, and internal markets simulate competitive signals. These mechanisms fall far short of the price system's spontaneous aggregation in a Hayekian sense, but they provide at least partial feedback loops that make coordination failures visible over time. For ethical knowledge, no comparable mechanism exists. No organizational process spontaneously aggregates the ethical implications of dispersed technical decisions into a signal that decision-makers can read. A caching decision that violates user privacy does not raise a price and a training-data choice that produces discriminatory outcomes does not register as a cost---at least not until the harm has materialized externally. The ethical domain is therefore not merely inadequately coordinated by organizational substitutes for the price system. Rather, it occupies a structural space in which no spontaneous aggregation mechanism operates at all.

This is why the ethical knowledge gap is more resistant to standard organizational fixes than knowledge problems in general: there is nothing to close it by default.

As Foss and Foss (2006) argue, authority can coexist with distributed knowledge, but only if the institutional design of rules, delegation, and control systems is carefully calibrated. Even then, the effectiveness of authority remains dependent on underlying information structures. Aghion and Tirole (1997) specify this point by distinguishing between \emph{formal authority}, which can be assigned by organizational fiat, and \emph{real authority}, which applies to those who actually possess the relevant information. This distinction is especially important for ethical knowledge in software development: ethical decision-making capacity cannot be understood solely in terms of formal hierarchy, but must be located where decision-relevant knowledge is actually available and can shape action.

\subsection{Application to Software Development}

Jensen and Meckling (1995) explore the tension between ``specific knowledge'' (held by employees) and ``general knowledge'' (held by managers) and how it affects firm structure. Analogously, in software production we may speak of the concrete distinction between \emph{implementation knowledge} (how a system is built and works) and \emph{application knowledge} (how and where a system will be used).

Consider the types of knowledge that reside with individual developers or small teams and are directly relevant to the ethical properties of a system. A developer knows that a particular input validation function silently drops malformed records rather than flagging them. Another knows that a caching layer retains user data for 72 hours after a deletion request is processed. A third knows that the machine learning model's training data was filtered in a way that systematically underrepresents a particular demographic group. A fourth knows that the fallback logic in the payment system rounds in the company's favor in edge cases.

Each of these is \emph{knowledge of the particular circumstances of time and place} in Hayek's sense. It is local to a specific point in the codebase, contextual to a specific design decision, and often tacit---the developer may not have a propositional representation of the knowledge, only a practical understanding of how the code behaves. Empirical studies of knowledge management in software development confirm this picture. Ryan and O'Connor (2013) found in a study of 46 software SMEs that tacit knowledge is acquired and shared primarily through quality social interactions rather than formal mechanisms, and that it predicts team effectiveness but not efficiency---suggesting that it matters most for the kind of work produced rather than the speed of production. Koskinen, Pihlanto, and Vanharanta (2003) showed that tacit knowledge in project work resists codification strategies and requires personalization---face-to-face interaction, proximity, shared experience. Boden et al.\ (2009) found that in distributed software teams, explicit knowledge can be shared through artifacts and tools, but tacit knowledge is strongly practice- and context-bound and requires social formats to become transportable. At the system level, MacCormack, Rusnak, and Baldwin (2012) provided empirical evidence for the `mirroring hypothesis', that states that software architectures reflect the communication structures of the organizations that build them thereby confirming that knowledge dispersal is not merely an individual-level phenomenon but is inscribed in the technical architecture itself.

The critical point is that each of these implementation details is potentially ethically significant. The silent data dropping may affect medical records. The caching behavior may violate privacy promises (or even regulations). The training data filtering may produce discriminatory outcomes. The rounding logic may constitute systematic financial harm. But whether the detail is ethically significant depends on contextual knowledge that the developer does not necessarily possess: who uses the system, under what conditions, with what vulnerabilities, in what regulatory environment and for what specific purpose.

Meanwhile, the organizational actors who possess this contextual knowledge lack access to the implementation details. The ethically relevant knowledge is, in the fullest Hayekian sense, distributed: it exists as what Fagin, Halpern, Moses, and Vardi (1995) call \emph{distributed knowledge}---the union of what all agents know may suffice for the ethical assessment, but no individual agent holds the union, and the organizational structure provides no mechanism for aggregating it.

A qualification is in order. The picture of sharp dispersal sketched above describes the idealized case, not every case as people obviously vary regarding their personal knowledge, motivation and experience. Experienced senior developers, for instance, often hold tighter bundles of implementation and contextual knowledge than the idealized separation suggests. A senior engineer who has been embedded in a domain for years has probably absorbed, through repeated cross-functional contact, fragments (small and large) of contextual knowledge that juniors never see: something about the customer base, the regulatory environment, the incidents that shaped current architectural decisions. For these developers, the gap between implementation knowledge and contextual knowledge is narrower, and in familiar decision contexts the two kinds of knowledge may even be sufficiently integrated to make ethically relevant relations visible without further aggregation. Yet, this integration is decision-specific rather than general: the same senior engineer entering an unfamiliar domain (a new product line, a new regulatory regime, a new user population) reverts to something closer to the junior's position---and these changes happen relatively often in software production compared to other industries. More importantly, where the integration does exist, it is idiosyncratic and personal rather than institutional. It lives in the individual's head, depends on their career history, motivation, and curiosity, and leaves the organization when they exit their role. A framework that relies on the tacit integration of senior developers is not an institutional solution to the dispersal problem but a band-aid. In fact, it is a gamble on individual virtue and the coincidence of a person's knowledge of a specific time and place. The dispersal claim should therefore be understood as describing the organizational baseline, meaning the condition an institutional design must be able to handle to address the structural knowledge problem an organization faces, regardless of whether particular individuals sometimes happen to bridge the gap in practice.

\subsection{Why Standard Aggregation Fails}

The natural organizational response is to create reporting structures: code reviews, architecture documents, risk assessments, compliance checklists. The Hayekian analysis suggests these face a fundamental limitation. Much of the relevant implementation knowledge is tacit. Even when articulable, the volume of potentially relevant detail is enormous so that no reporting system can capture all of it.

Furthermore, the developer does not know which details to report, precisely because they lack the contextual knowledge that would make the ethical relevance visible. An instruction to ``report anything with ethical implications'' presupposes the very knowledge the reporting is supposed to aggregate in the first place. Cataldo et al.\ (2006) have shown empirically that coordination requirements in software projects are highly volatile and frequently extend beyond team boundaries. When actual coordination activities match these emergent requirements, development time decreases significantly. The implication for ethical coordination is direct: the knowledge about ``who needs to talk to whom about what'' arises locally, situationally, and changes rapidly, which is a precise organizational analogue of Hayek's dynamic knowledge fragments.

%---------------------------------------------------------------------
\section{Organizational Epistemic Failures: Interpretive Deficit and Credibility Attenuation}
%---------------------------------------------------------------------

The Hayekian analysis establishes that ethically relevant knowledge is dispersed and resists centralization. If this were the only problem, the standard Hayekian prescription, which has been to push decision authority to local agents, might suffice. But two further mechanisms ensure that decentralization alone cannot close the ethical knowledge gap. Organizations fail both to render ethical saliences intelligible and to route relevant knowledge to the people who can act on it. The first failure is an \emph{interpretive deficit}, analyzable through the sensemaking tradition in organizational theory. The second is a \emph{credibility attenuation}, which needs to be looked at through the lens of social epistemology of testimony and epistemic dependence.

\subsection{Interpretive Deficits as a Structural Epistemic Problem in Software Development}

The Hayekian problem of dispersed knowledge is not exhausted by the fact of dispersion itself. Even if ethically relevant knowledge is present somewhere in the organization, it may fail to matter in practice unless it is made intelligible within the interpretive framework of those making local decisions. The question is therefore not only \emph{who} possesses relevant knowledge, but \emph{how} such knowledge becomes visible as relevant at all. This shifts the analysis from the distribution of knowledge to the organizational processes through which situations are made sense of. While Hayek discusses the dispersal of knowledge in societies, Hutchins (1995) provides a micro-level view of this phenomenon within a technical team. To navigate a massive ship in narrow waters, the crew must constantly determine the ship's precise position. This is done through a process called ``fixing'' the position. Just as no single member of a ship's navigation crew possesses the `whole' knowledge of the ship's position, no single developer or product manager possesses the full ethical picture of a software system. The knowledge is distributed: spread across Jira tickets, lines of code, different specialized roles, and application domain knowledge.

Weick's (1995) sensemaking framework conceives organizational meaning-making as a practical, ongoing, and fundamentally social process. Organizations do not passively receive and process information. Rather they enact environments by selecting certain cues, interpreting them through available categories, and acting on the basis of what thereby becomes intelligible. Sensemaking is retrospective, socially stabilized, and oriented toward plausibility rather than truth in any strict representational sense: actors settle on interpretations that are workable within existing frames rather than on exhaustive or maximally accurate descriptions of reality (Weick 1995; Weick, Sutcliffe, and Obstfeld 2005).

This insight carries a broader epistemic implication: interpretation is not secondary to judgment but one of its conditions of possibility. Before an issue can be evaluated, weighed, or translated into a requirement, it must first be rendered intelligible as the kind of thing that calls for attention at all. Ethical deliberation, in particular, presupposes prior legibility. A team cannot meaningfully assess whether a design choice is fair, harmful, or responsible if that choice is not first understood under a description to which such presumptions apply. Failures at the level of interpretation therefore propagate downstream into failures of deliberation and decision-making. What is not interpretively available does not enter reasoning simply as a rejected consideration. It fails to arise as a consideration at all.

The interpretive deficit in software development can be understood precisely in these terms. Engineering practice provides a highly refined sensemaking apparatus for rendering technical work intelligible. It offers rich and suitable categories for reasoning about architecture, efficiency, scalability, maintainability, optimization and reliability. At the same time, it institutionalizes boundaries---most notably between ``technical'' and ``non-technical'' concerns---that structure professional identity and responsibility. These boundaries are not merely descriptive but performative. They guide attention, define relevance, and limit what counts as a legitimate problem within the engineering role. In this respect, they function as structured frames or institutional logics that enable coordinated action while simultaneously constraining what can be perceived and articulated (Kaplan 2008; Thornton, Ocasio, and Lounsbury 2012). As outlined above, one might expect seniority to mitigate the interpretive deficit. Developers with deeper experience have, after all, been exposed to more contexts and seen more of how their technical choices play out. But the sensemaking literature suggests a more complicated picture. Expertise deepens mastery of the interpretive categories of a practice. It does not automatically generate new categories from outside that practice. A senior developer has spent a decade refining his or her ability to classify decisions as architectural, performance-related, or maintenance-related. Thus, their engineering frame is not merely available but deeply entrenched. Weick (1995) has described this phenomenon as \emph{enactment lock-in}: the richness of a well-practiced frame may make alternative framings harder to access, not easier, because every new situation is routed through the familiar categories before alternative readings become salient. The senior developer who knows both that the system is used by vulnerable populations and how the caching layer behaves may still encounter the caching decision as a performance trade-off, because the two bodies of knowledge sit in separate compartments within the engineering frame rather than being interpretively integrated. Seniority thus may weaken the dispersal mechanism for some decisions while potentially reinforcing the interpretive deficit for others. This is an empirical claim that warrants investigation, but it suggests that the prima facie expectation, which is that experienced developers will naturally see ethical issues that juniors may miss, underestimates the structural aspect of professional sensemaking.

From this perspective, ironically, the interpretive deficit is not a breakdown of sensemaking but a consequence of its very success within a restricted frame. Developers routinely make competent and well-justified decisions within the categories available to them. A caching decision, for example, is readily interpreted as a trade-off between storage cost and performance. Within the engineering frame, this is a complete and satisfactory description of the problem. Because it is satisfactory, there is no trigger to search for alternative interpretations. The ethical dimension is not actively dismissed. Rather, it never gains salience as a relevant dimension of the situation.

Another example is the following: A developer introduces detailed logging in order to be able to debug system behavior. Within the technical interpretive frame, this is a standard and prudent practice for observability and maintenance as it allows to track system behavior linked to specific actions. What does not enter the interpretation is that logs may contain sensitive user behavior data that can later be repurposed or misused. The decision is understood as improving system reliability, not as expanding surveillance capabilities which---at a later point---can lead to violations of user privacy.

This phenomenon is well documented in organizational and science and technology studies. Research on framing and categorization shows that decision-makers attend selectively to aspects of a situation that fit established categories, while other aspects remain effectively invisible (Bowker and Star 1999; Kaplan 2008). Studies of distributed cognition and epistemic cultures demonstrate that knowledge in complex systems is not located in individual minds but distributed across roles, artifacts, and practices, such that no single actor has access to the full meaning of a situation (Hutchins 1995; Knorr Cetina 1999). What counts as a relevant feature is therefore always mediated by local practices and tools. In such settings, interpretive gaps are not anomalies but routine consequences of how knowledge is organized.

Software engineering amplifies these general dynamics. Its practices are built around abstraction, modularization, and division of labor. Developers typically engage with systems through local artifacts---tickets, modules, interfaces, and metrics---rather than through the full social context of use. The ethically relevant significance of a decision, however, often emerges only at the level of system integration, organizational deployment, or actual user experience. This creates a structural misalignment between the point of decision and the location of consequence. The cues that would render a decision ethically salient are therefore often distributed across different roles, different artifacts, and different temporal stages.

Empirical research in software engineering and related fields supports this diagnosis. Studies of requirements engineering show that crucial contextual knowledge about users and environments of use is frequently incomplete, tacit, or weakly integrated into development processes (Nuseibeh and Easterbrook 2000). Ethnographic work on software teams documents how developers rely on locally available representations and often lack visibility into broader system impacts (de Souza 2005). Research in HCI and responsible AI similarly finds that ethical concerns---such as fairness, privacy, or social impact---are often recognized in principle but fail to translate into concrete design and implementation decisions due to missing conceptual and organizational linkages (Madaio et al.\ 2020; Rakova et al.\ 2021). Practitioners frequently report knowing that ethical issues exist while lacking actionable ways to connect them to their everyday technical work.

These findings reinforce that the interpretive deficit is not adequately described as only a lack of information. Developers may possess many of the relevant facts. They may know that a system processes sensitive data, affects vulnerable populations, or operates in high-stakes domains. What is missing is not awareness in the abstract, but a pathway of translation from contextual knowledge to local decision-making. Organizational language regimes and professional vocabularies determine which connections can be readily made and which remain difficult to articulate (Wilmot 2024). Where no established pathway links a technical implementation detail to its broader social significance, that significance remains interpretively inert.

The interpretive deficit can therefore be understood as a specific form of organizational epistemic limitation. It arises when actors are equipped with sophisticated tools for solving technically defined problems but lack equally developed resources for recognizing how those problems are embedded in wider social and normative contexts. Crucially, this limitation does not depend on individual negligence or lack of concern. It is structurally supported by the way knowledge, attention, and responsibility are organized in software development. The result is a systematic asymmetry: technical features of a situation are rendered highly visible and actionable, while ethical features remain latent, dispersed, and difficult to bring into view.

Understood in this way, the ethical knowledge gap is, in part, an interpretive gap. It consists in the absence of sufficiently rich categories, vocabularies, and practices through which technically intelligible decisions can also become ethically intelligible. Addressing this gap therefore requires more than additional information or ex post evaluation. It requires interventions at the level of sensemaking itself: expanding the interpretive resources through which developers understand their work, and creating organizational conditions under which ethically relevant aspects of technical decisions can become visible within the very process by which those decisions are made.

\subsection{Epistemic Dependence and Credibility Attenuation}

Even when a developer overcomes the interpretive deficit---when they recognize that a technical decision has ethical implications and attempt to communicate this---a second mechanism may prevent the insight from shaping collective judgment. Modern organizations, like modern knowledge more generally, are structured by deep epistemic dependence. In settings marked by specialization and division of labor, actors must routinely rely on the testimony of others whose underlying evidence they cannot themselves fully inspect or evaluate (Hardwig 1985). This is not a bug, but a constitutive and unavoidable feature of complex knowledge practices. As the literature on expertise emphasizes, the problem is not whether organizations should depend on others' knowledge, but how they should distribute epistemic authority: who is treated as a credible speaker on which kinds of claims (Goldman 2001; Zagzebski 2012).

Organizations usually solve this problem through role-based allocations of credibility. Developers are treated as authoritative on implementation and system behavior, product managers on markets and users, compliance officers on regulation, and---where present---ethics specialists on explicitly normative questions. This partitioning of authority is broadly rational. It reduces the burden of evaluating every claim from scratch and provides workable heuristics for coordinating expertise under conditions in which no single actor can know enough to judge everything independently (Goldman 2001; Zagzebski 2012). Yet the same structure that makes epistemic dependence manageable also generates a systematic difficulty when relevant observations do not fit neatly within these established domains.

This is especially important in software development because many ethically significant observations are hybrid in form. They combine implementation-level knowledge with contextual or normative judgment. When a developer says that a caching design creates a privacy risk for vulnerable users, the force of the claim lies precisely in this conjunction: the speaker possesses detailed knowledge of the technical system and enough contextual understanding to recognize the broader significance of that design choice. But organizational credibility structures are often unable to receive such claims in their integrated form. The developer is heard as competent on the technical aspect of the issue, but not necessarily as authoritative on its ethical significance. As a result, the claim does not disappear entirely, but its epistemic force is weakened as it moves through organizational channels.

This weakening can be described as \emph{credibility attenuation}. By this I mean that a relevant observation is not simply ignored, but loses epistemic efficacy as it passes through organizational filters. The research report identifies several recurrent forms of attenuation. First, the claim may be \emph{reframed}: an ethical warning is translated into a purely technical issue, so that the speaker's recognized expertise is preserved while the normative dimension is stripped away. Second, it may be \emph{deferred} or routed to the ``appropriate'' authority, e.g.\ privacy, compliance, legal, or policy, without preserving the implementation-level reasons that originally made the issue visible. Third, it may be treated as \emph{overreach}, that is, dismissed on the grounds that the speaker is operating outside their authorized domain. Finally and potentially worst, attenuation may take the form of \emph{silencing}, where anticipated non-uptake or sanction discourages agents from raising similar concerns in the first place.

The distinction between reliance and trust sharpens this point. Faulkner (2011) argues that testimonial knowledge cannot be understood merely in terms of predictive reliance---expecting someone to perform reliably based on past behavior---but involves trust in a thicker sense: a normative orientation in which another's contribution is awarded genuine and benevolent uptake within a shared epistemic practice (Faulkner 2011). Applied to organizations, this suggests that companies often \emph{rely} on developers to produce functioning code, resolve bugs, and optimize systems, but do not necessarily \emph{trust} them as speakers on the ethical significance of their own technical choices. Their claims may be heard, but not as contributions that decision-makers are expected to treat as reasons within collective deliberation. In this respect, credibility attenuation is not only a matter of communication failure, but of incomplete epistemic recognition.

This problem is reinforced by organizational sensemaking processes as discussed above. As Weick, Sutcliffe, and Obstfeld (2005) and Maitlis and Christianson (2014) show, organizations do not simply transmit claims neutrally. They rather actively shape what counts as a plausible description of a situation and who is entitled to define that situation. Uptake is therefore not binary. Claims are translated, reformulated, and fitted into existing organizational vocabularies. In role-differentiated organizations, this means that cross-domain observations are especially likely to be transformed into forms that align with pre-existing authority boundaries. The ethical concern remains present only in attenuated form, while the original integration of technical detail and contextual judgment is lost.

Research on organizational silence and psychological safety adds an important empirical layer. Edmondson's work on psychological safety shows that individuals are more likely to raise concerns, ask questions, and challenge assumptions when they expect it to be interpersonally safe to do so (Edmondson 1999). Morrison and Milliken (2000), by contrast, show how organizations can produce silence as a collective pattern: employees withhold information about potential problems because they expect speaking up to be ineffective, risky, or unwelcome. In the present context, this means that credibility attenuation does not merely affect whether a voiced concern is taken up. It also feeds back into whether such concerns are voiced in the first place. When ethical observations emerging from technical roles are repeatedly reframed, routed away, or dismissed, silence, in a way, becomes a rational organizational adaptation.

The role of hierarchy further supports this analysis. Kwok's (2021) account of workplace hierarchy shows that organizational status structures constrain the range of topics on which agents are treated as credible speakers. Knowledge generated at the level of production may fail to count as knowledge once it crosses status or authority boundaries. Even if one sets aside the stronger normative vocabulary of epistemic injustice (c.f.\ Fricker 2007) used in Kwok (2021), the core analytical point remains highly relevant: hierarchical workplaces do not merely distribute tasks, but also distribute credibility. In software organizations, this is particularly consequential because the actors with the most detailed knowledge of implementation are rarely the same actors who possess formal authority to define product direction, risk significance, or organizational priorities.

Empirical studies in software and AI practice make this mechanism especially visible. Holstein et al.\ (2019) show that practitioners working on fairness in machine learning encounter barriers that are not merely technical, but organizational: unclear responsibility allocation, limited decision rights, and misalignment between ethical concerns and existing workflows. Madaio et al.\ (2020) similarly find that fairness work often depends on ad hoc advocacy by motivated individuals and becomes durable only when supported by organizational infrastructures such as checklists, routines, and aligned processes. Widder and Nafus (2023) add an especially important software-specific finding: modularity and ``AI supply chain'' thinking encourage engineers to locate responsibility elsewhere---upstream, downstream, or in another team---so that harms are recognized in principle but systematically displaced in practice. These studies suggest that software organizations frequently do not lack awareness altogether. Rather, they fail to preserve and integrate ethically relevant observations when these emerge from local technical positions.

Understood in this way, credibility attenuation is the second half of the ethical knowledge gap. Hayekian knowledge dispersion explains why relevant knowledge is distributed across actors and contexts rather than concentrated in a single point. The interpretive deficit explains why ethically relevant features may fail to become visible within local frames of action. Credibility attenuation explains why, even when such features become visible and are articulated, organizations may still fail to act on them because their structures of epistemic authority cannot adequately receive hybrid judgments. The problem, then, is not only that ethical knowledge is dispersed, nor only that it is difficult to interpret, but also that it is insufficiently taken up when it appears in forms that cross established role boundaries.

\begin{figure}[htbp]
\centering
\begin{tikzpicture}[
    font=\sffamily\footnotesize,
    >=Latex,
    node distance=6mm and 8mm,
    box/.style={
        rectangle,
        rounded corners=2pt,
        draw=blue!55!black,
        thick,
        fill=blue!3,
        text width=5.6cm,
        minimum height=8mm,
        align=center,
        inner sep=4pt
    },
    narrowbox/.style={
        box,
        text width=4.8cm
    },
    arrow/.style={
        -{Latex[length=2.2mm,width=1.6mm]},
        thick,
        draw=gray!70!black
    }
]
% Top: structural precondition
\node[box] (A) {Epistemic Dependence\\{\scriptsize specialized division of intellectual labor}};
\node[box, below=of A] (B) {Organizational Allocation of Epistemic Authority\\{\scriptsize role- and hierarchy-based}};

% Middle: four attenuation forms in 2x2 grid
\node[narrowbox, below left=9mm and 1mm of B] (C1) {Reframing\\{\scriptsize ethical $\rightarrow$ technical issue}};
\node[narrowbox, below right=9mm and 1mm of B] (C2) {Deferral / Routing\\{\scriptsize passed on without reintegration}};
\node[narrowbox, below=of C1] (C3) {Dismissal as Overreach\\{\scriptsize speaker outside their domain}};
\node[narrowbox, below=of C2] (C4) {Anticipated Attenuation\\{\scriptsize raising concerns becomes costly}};

% Consequences
\node[box, below=10mm of C3] (D) {Loss of Integrated Judgment\\{\scriptsize implementation knowledge $+$ ethical significance}};
\node[box, below=10mm of C4] (E) {Organizational Silence\\{\scriptsize reduced cross-boundary learning}};

% Outcome
% Outcome
\node[box, below=16mm of $(D)!0.5!(E)$] (F) {Outcomes: delayed or absent mitigation,\\ethical blind spots, coordination failures};

% Main flow
\draw[arrow] (A) -- (B);
\draw[arrow] (B.south) -- (C1.north);
\draw[arrow] (B.south) -- (C2.north);
\draw[arrow] (C1) -- (C3);
\draw[arrow] (C2) -- (C4);
\draw[arrow] (C3) -- (D);
\draw[arrow] (C4) -- (E);
\draw[arrow] (D.south) -- (F.north west);
\draw[arrow] (E.south) -- (F.north east);
\end{tikzpicture}
\caption{Forms of credibility attenuation}
\label{fig:credibility-attenuation}

\end{figure}
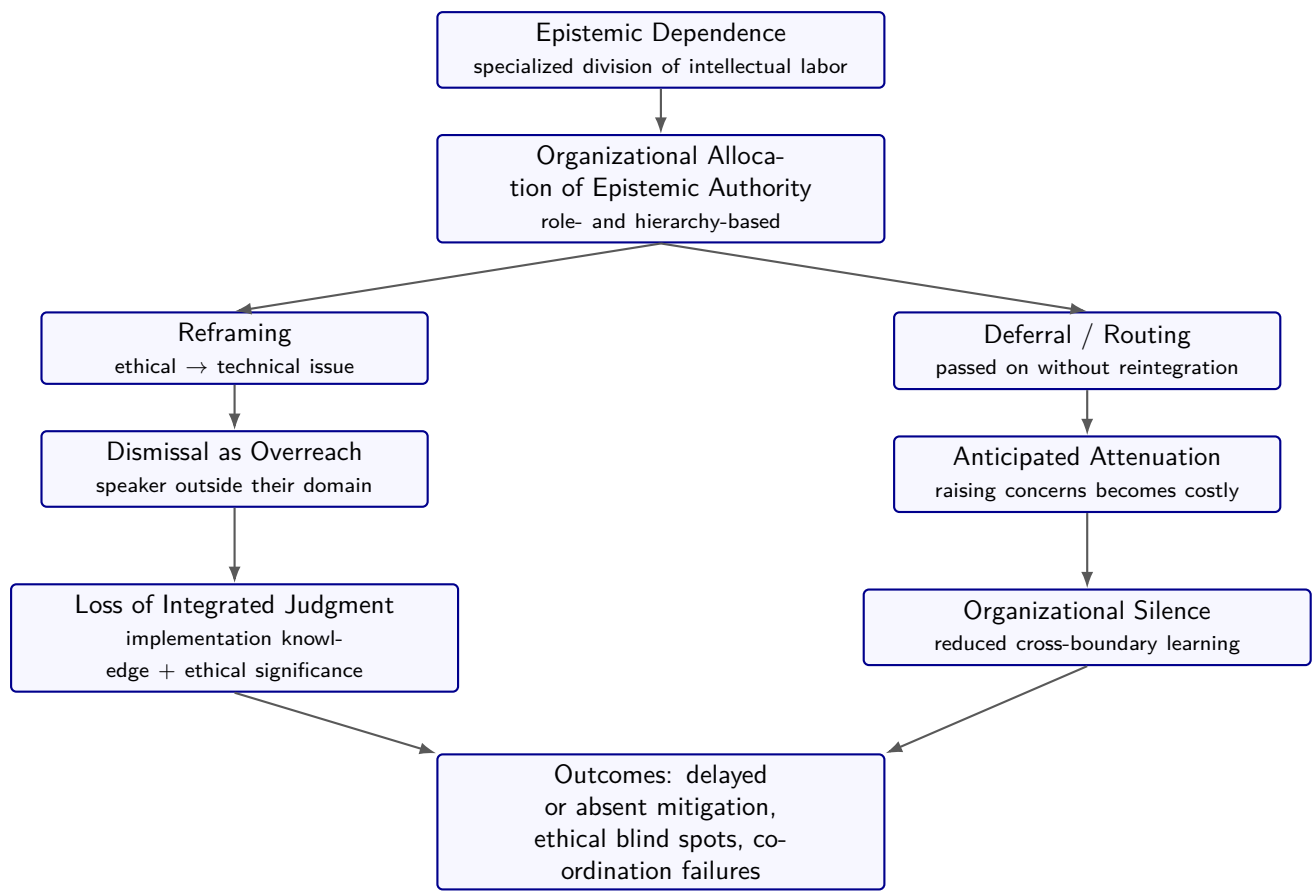

\subsection{The Structural Character of Both Mechanisms}

It is worth noting that both mechanisms are structural, not merely individual. The interpretive deficit is produced by the sensemaking apparatus of engineering culture and the categorical structure of organizational language, not by any developer's personal failure to think ethically. The credibility attenuation is produced by the epistemic-dependence architecture that organizations use to manage the division of intellectual labor, not by any manager's personal disregard for developer input. Both persist even when all individual actors are well-intentioned, competent, and ethically attentive, albeit probably to a lesser degree.

This structural character means the problem cannot be solved by individual virtue or effort alone. Ethics training may expand individual interpretive resources, but unless it restructures the organizational sensemaking categories through which decisions are classified, the interpretive deficit reconstitutes itself in practice. Encouraging developers to ``speak up'' may motivate more cross-boundary communication, but unless the epistemic-dependence architecture changes---unless the organization begins to \emph{trust} and to take seriously developers as ethical speakers, not merely \emph{rely} on them as technical producers, those attempts will be systematically attenuated. The problem is institutional, and it thus requires an institutional response.

%---------------------------------------------------------------------
\section{The Combined Framework: The Ethical Knowledge Gap}
%---------------------------------------------------------------------

\subsection{The Tripartite Blockage}

The Hayekian layer and the two organizational epistemic failures interact to produce what I call the \emph{ethical knowledge gap}: a structural condition in which the knowledge required for ethically informed decision-making is systematically unavailable at the point of decision. The gap has three components, and it is their interaction---not any single component---that makes the problem resistant to standard interventions.

The first component is \emph{dispersal}: ethically relevant knowledge is distributed across multiple agents and positions, is in significant part tacit, and cannot be fully centralized. The second is the \emph{interpretive deficit}: the sensemaking apparatus available to the agents who hold implementation knowledge does not provide the categories needed to recognize its ethical significance---the knowledge is present but ethically inert. The third is \emph{credibility attenuation}: where a developer does perceive and articulate an ethical concern, the organization's epistemic-dependence structure discounts the testimony as it crosses role boundaries.

The three components create a trap. Centralization fails because the knowledge is tacit and dispersed (Hayek). Decentralization fails because the local sensemaking apparatus cannot render ethical significance visible (interpretive deficit). Hybrid approaches that encourage bottom-up reporting fail because the epistemic-dependence architecture attenuates cross-boundary testimony (credibility attenuation). No single-vector intervention can close the gap because the gap is sustained by three independent mechanisms.

To make the interaction of the three mechanisms concrete, consider a schematic but realistic scenario: A data engineering team builds an automated pipeline for an insurance company. The pipeline aggregates customer data from multiple sources, e.g.\ purchase history, geographic data, credit scores, to generate risk profiles used in pricing decisions. A data engineer notices during development that the pipeline includes a geocoding step that maps customer addresses to census-block-level demographic data, effectively encoding neighborhood racial and socioeconomic composition into the risk score without using race as an explicit variable.

The dispersal mechanism is immediately visible. The data engineer knows the pipeline architecture. Specifically, that the geocoding step feeds census-block features into the model, and that these features correlate strongly with race and income at the neighborhood level. The actuarial team, that manages financial risks for insurance companies, knows that the risk model is used for health and life insurance pricing. The legal department knows that using race as a rating factor is prohibited in many jurisdictions but that proxy variables occupy a legal gray zone. The marketing team knows that the company publicly positions itself as committed to ``AI-driven fairness'' in pricing. The management knows that for ethical and strategic resp.\ reputational reasons the use of proxy variables should be avoided. The ethically relevant relation that the pipeline encodes racial proxies into insurance pricing in a way that may constitute automated redlining while the company publicly claims fairness, requires the conjunction of all five knowledge fragments. Yet, no single actor holds the full picture.

The interpretive deficit determines how the data engineer encounters the geocoding step. Within the engineering sensemaking frame, census-block data is a data enrichment problem: it adds predictive power to the model, and that is why it is there. The engineer's working categories are precision, recall, feature importance, and data quality---not disparate impact, proxy discrimination, or redlining. The engineer may even know, in the abstract, that geographic features can correlate with demographic variables. But the interpretive pathway from ``this feature correlates with demographics'' to ``we are building an automated redlining system'' is not readily available in the data engineering vocabulary. The technical interpretation (``geographic features show high predictive value'') is plausible, complete, and satisfying within the engineering frame. There is no inherent trigger to seek an alternative reading.

The credibility attenuation operates if the engineer does connect the dots and raises the concern (``The census-block features are basically encoding race or poverty into the risk score.''). The organizational response follows characteristic patterns. The concern may be technically reframed (``Can you run a disparate impact analysis?''), converting a structural design concern into a measurement task. It may also be deferred (``That sounds like a compliance question; let's put legal in the loop''). Yet legal, lacking understanding of how the geocoding step actually works, issues a generic advisory to ``proceed with caution'' that has no operational meaning for the engineering team. Or it may be displaced through modularity (``We just build the model. The actuarial team decides how to use it'') so that each team can point to another as the locus of responsibility while the system as a whole produces discriminatory outcomes. This last form illustrates what Widder and Nafus (2023) have documented as the displacement of accountability through AI supply chain thinking: modular system architecture maps onto modular moral responsibility, and the integrated ethical observation disintegrates as it passes through organizational boundaries.

It is worth noting that the ethical knowledge gap is symmetric. The same structural conditions that prevent organizations from recognizing ethical risks also prevent them from recognizing ethical opportunities. Where implementation knowledge and contextual knowledge fail to meet, teams may fail not only to notice harms but also to notice where the software could better support users, respect domain values, or enable forms of interaction that would more fully realize the good the technology is capable of producing. A developer who understands the fine-grained behavior of a notification system but not the clinical context of its users is as unable to see how the system might genuinely support recovery as they are to see how it might cause distress. A product owner who knows that the platform serves job seekers with care responsibilities (e.g.\ looking after children etc.), but does not know how the ranking features are computed, cannot see either side of the opportunity: they cannot recognize that the current system penalizes career gaps, and they cannot imagine how a differently designed system might treat such gaps as meaningful signals of care work, resilience, or life experience worth surfacing to employers. The framework is therefore not only a diagnostic of ethical failure but of ethical underperformance more generally: it identifies the structural conditions under which software falls short of what it could be, whether the shortfall takes the form of harm produced or of good left unrealized.

\subsection{A Formal Sketch}

I offer a schematic formalization of the ethical knowledge gap. The aim is not to provide a fully operational model, but to make the layered structure of the argument explicit. The gap arises from the interaction of three mechanisms: dispersed access to relevant facts, failure to recognize their ethical significance, and attenuated uptake when concerns are communicated across organizational boundaries.

Let $D$ be the set of development decisions, and let $d \in D$ denote a particular decision. For each $d$, let
\[
I(d)
\]
be the set of implementation facts relevant to $d$, and
\[
X(d)
\]
the set of contextual facts relevant to $d$. Since ethical significance usually arises not from either class of facts in isolation but from their relation, define
\[
R(d) \subseteq I(d) \times X(d)
\]
as the set of ethically relevant relations for decision $d$.

Let $A$ be the set of relevant organizational agents, and for each $a \in A$, let
\[
A_a(d) \subseteq I(d) \cup X(d)
\]
be the set of facts accessible to agent $a$ with respect to decision $d$. This captures the dispersed-knowledge structure: the organization as a whole may collectively contain the relevant facts,
\[
\bigcup_{a \in A} A_a(d) \supseteq I(d) \cup X(d),
\]
while no individual agent normally possesses the whole picture,
\[
A_a(d) \subsetneq I(d) \cup X(d).
\]

To capture which ethical issues are even visible under this distribution of access, define
\[
V(d) = \bigl\{ r \in R(d) \mid \exists\, a \in A \text{ such that the facts needed for } r \text{ are contained in } A_a(d) \bigr\}.
\]
$V(d)$ is the set of ethically relevant relations visible somewhere in the organization.

For each agent $a$, let
\[
V_a(d) = \bigl\{ r \in V(d) \mid \text{the facts needed for } r \text{ are contained in } A_a(d) \bigr\}
\]
be the set of ethically relevant relations visible in principle to that agent, and let
\[
L_a(d) \subseteq V_a(d)
\]
be the set of ethically relevant relations that agent $a$ actually recognizes. The hermeneutical deficit of agent $a$ can then be expressed as
\[
H_a(d) = 1 - \frac{|L_a(d)|}{|V_a(d)|} \quad \text{for } |V_a(d)| > 0.
\]
This measures the share of ethically relevant relations visible to agent $a$ that remain unrecognized.

Recognition alone, however, is insufficient. Let $m_r$ denote a message communicating ethically relevant relation $r$, and define
\[
\tau(a, b, m_r) \in [0, 1]
\]
as the uptake coefficient of that message when sent from agent $a$ to agent $b$. Let
\[
\theta \in (0, 1]
\]
be the minimum threshold required for a communicated concern to count as institutionally effective. We can then define
\[
U(d) = \bigl\{ r \in R(d) \mid \exists\, a, b \in A \text{ such that } r \in L_a(d) \text{ and } \tau(a, b, m_r) \geq \theta \bigr\}
\]
as the set of ethically relevant relations that become actionable at the point of decision.

Finally, the ethical knowledge gap for decision $d$ is
\[
G(d) = 1 - \frac{|U(d)|}{|R(d)|} \quad \text{for } |R(d)| > 0.
\]
$G(d)$ is thus the proportion of ethically relevant relations that fail to become actionable. The model makes explicit that the gap is cumulative: dispersal limits what is visible at all, hermeneutical deficit limits what is recognized, and attenuated uptake limits what enters decision-making. The ethical knowledge gap is the resulting shortfall between the ethical structure objectively present in a decision and the ethical structure the organization can actually bring to bear on it.

\subsection{The Structural Blockage Claim}

The formal sketch highlights the structural blockage claim: absent institutional mechanisms that simultaneously address dispersal, interpretive deficit, and credibility attenuation, ethical software development is not merely difficult but structurally blocked. The organizational structures responsible for producing software---the division of labor, the sensemaking categories, the epistemic-dependence architecture---are constitutively at odds with the epistemic conditions for ethical decision-making. The gap is not a contingent failure of any particular organization, but rather a structural feature of how software development is organized as a practice.

Suppose a single decision $d$ contains ten ethically relevant relations, so that $|R(d)| = 10$. Because relevant facts are dispersed across roles, only six of these relations are visible somewhere in the organization, so $|V(d)| = 6$. Because interpretive resources are limited, only three are actually recognized, so $|\bigcup_{a \in A} L_a(d)| = 3$. Of these, only one is taken up strongly enough to shape the decision, so $|U(d)| = 1$. The ethical knowledge gap is therefore
\[
G(d) = 1 - \frac{|U(d)|}{|R(d)|} = 1 - \frac{1}{10} = 0.9.
\]
In this case, ninety percent of the ethically relevant structure of the decision fails to become actionable.

This illustrates precisely why the ethical knowledge gap is not a simple information deficit. It is produced sequentially by dispersed knowledge, hermeneutical failure, and credibility attenuation.

%---------------------------------------------------------------------
\section{Implications for Institutional Design}
%---------------------------------------------------------------------

\subsection{The Need for a Translation Layer}

If the ethical knowledge gap is sustained by three independent mechanisms, an adequate institutional response must address all three simultaneously. What is needed is a \emph{translation layer}---an institutional mechanism that mediates between the dispersed knowledge of developers and the contextual knowledge of organizational decision-makers, while counteracting the interpretive deficit and credibility attenuation. It differs from participatory approaches such as value-sensitive design (cf.\ Section~2.3) in a specific respect: the translation layer does not merely assemble actors with different knowledge but intervenes in the conditions under which their knowledge becomes ethically legible and credibly communicable in the first place.

A translation layer is not a communication channel. It is a structured institutional practice that: (1)~enriches the sensemaking categories available to developers, introducing interpretive resources that make the ethical dimension of technical decisions visible (expanding $L_a(d)$ toward $R_a^*(d)$). (2)~restructures the epistemic-dependence architecture by creating spaces where developers are trusted---not merely relied upon---as speakers on the ethical implications of their technical work (increasing $\tau$); and (3)~provides structured mechanisms for aggregating tacit, local knowledge without requiring full centralization (addressing Dispersal through targeted rather than comprehensive aggregation).

A three-phase model of ethical software development (Gogoll and Zuber 2026)---identification, deliberation, implementation---maps onto this structure, though not as a one-to-one correspondence between phases and mechanisms. Identification addresses both dispersal and the interpretive deficit simultaneously: it surfaces contextual knowledge held by stakeholders, domain experts etc. bringing it into contact with the implementation knowledge held by developers, while at the same time introducing new sensemaking categories through which this combined knowledge can be read as ethically significant. The two functions are inseparable in practice, because surfacing contextual knowledge without the interpretive categories to integrate it with implementation detail would leave the information inert, while expanding sensemaking categories without contextual knowledge to apply them to would leave them empty. Deliberation addresses credibility attenuation: it creates structured epistemic spaces where the normal organizational epistemic-dependence hierarchy is suspended and technical actors' observations about ethical implications receive full uptake. Implementation, again, addresses dispersal: rather than centralizing all ethically relevant knowledge, it distributes ethical constraints---translated into technical specifications, tests, and guardrails---back to the local level where they can be acted upon with the relevant tacit knowledge.

The translation layer is specified here at the level of functional requirements rather than institutional prescription. This is not a limitation but a consequence of the framework itself. A fully specified, universally applicable institutional design would presuppose precisely the kind of centralizable knowledge that the Hayekian layer of the argument shows to be unavailable. The concrete form of the translation layer must be determined locally, in light of the specific knowledge structures, sensemaking cultures, and epistemic-dependence architectures of each organization looked at individually. What the framework provides is thus not a blueprint but a set of conditions: any institutional mechanism that claims to address the ethical knowledge gap can be evaluated against whether it simultaneously counteracts dispersal, the interpretive deficit, and credibility attenuation.

\subsection{Design Principles}

If the ethical knowledge gap is produced by three independent mechanisms, design principles for a translation layer must be derived from, and mapped onto, those mechanisms. The following principles are therefore not freestanding recommendations but structural responses to the specific blockages identified in the preceding analysis.

\textbf{Bidirectional knowledge flow} (addressing dispersal). The translation layer must facilitate the flow of contextual knowledge to developers (enabling ethical recognition) and implementation knowledge from developers (enabling contextual actors to perceive technical risks). This directly targets the Hayekian dispersal mechanism: because ethically relevant relations in $R(d)$ require elements from both $I(d)$ and $X(d)$, any mechanism that moves knowledge in only one direction leaves the dual nature of the cases unaddressed.

\textbf{Sensemaking enrichment, not mere information provision} (addressing the interpretive deficit). Providing developers with information about use contexts is insufficient if the sensemaking categories remain unchanged. What is needed is not a briefing but a new vocabulary---a restructuring of the categories through which developers enact their working environment. This is a sensemaking intervention in Weick's sense: it changes not what developers know but how they make sense of what they know. In the insurance pipeline example from Section~5.1, the data engineer possessed the technical fact that census-block features correlate with demographics. What was missing was not information but an interpretive pathway from ``geographic features show high predictive value'' to ``we are building an automated redlining system.'' Sensemaking enrichment installs precisely such pathways.

\textbf{From reliance to trust} (addressing credibility attenuation). Telling developers to ``speak up'' without changing the epistemic-dependence architecture may turn out to be futile. The organization must move from merely \emph{relying} on developers for technical output to \emph{trusting} them as speakers whose ethical observations deserve uptake. This requires structural intervention at the level of organizational roles and decision procedures---for example, structured deliberation formats where cross-domain testimony is a required input rather than an optional contribution. This requires the spending of resources---especially time.

\textbf{Targeted aggregation over comprehensive centralization} (addressing dispersal). Full centralization is impossible (Hayek), but the aspiration to identify the ethical implications of a decision is not thereby misguided---it is the mode of pursuit that must change. The comprehensive question ``What are all the ethical implications?'' remains the normative horizon of ethical software development. But the organizational mechanism for approaching it must also work through targeted, local aggregation rather than centralized collection alone: not only to gather all relevant knowledge at one point, but to expand the accessible knowledge set $A_a(d)$ at identified risk points through structured local inquiry. The comprehensive question sets the direction, while targeted aggregation is the epistemically feasible way of moving toward it. Jensen and Meckling's (1995) framework of specific knowledge and decision rights provides the organizational-design vocabulary; ethical decision rights should be co-located with the specific knowledge they require, supplemented by control mechanisms.

\textbf{Integrated design} (addressing the interaction of mechanisms). The paper's central claim is that the three mechanisms are independently sufficient to sustain the ethical knowledge gap: closing one while leaving the others intact produces no net improvement. This imposes a meta-principle on institutional design: interventions must be evaluated not individually but as configurations that address dispersal, interpretive deficit, and credibility attenuation jointly. An organization that enriches sensemaking categories but does not reform its epistemic-dependence architecture will produce developers who see ethical issues but whose testimony is attenuated. An organization that reforms credibility structures but does not address sensemaking will empower developers to speak---about concerns they cannot recognize. The design challenge is therefore combinatorial, not additive.

\subsection{Differentiating the Translation Layer by Value Type}

The ethical knowledge gap is not uniform across value types, and the framework's analytical precision can be sharpened by distinguishing two kinds of ethical value in software development as proposed by Gogoll and Zuber (2026). \emph{Techno-generic values} are concerns that arise from the nature of the technology itself regardless of deployment context---fairness and bias in machine learning, transparency in automated decision-making, robustness in autonomous systems. \emph{Domain-specific values} are tied to the sector in which the software is deployed: healing in healthcare, student agency in education, proportionality in law enforcement. The same technology can carry very different ethical weight depending on the domain. This distinction maps directly onto the formal structure. Techno-generic values can in principle be derived from $I(d)$ alone: if you are building an ML classifier, fairness concerns arise from the statistical properties of the artifact regardless of context. Domain-specific values, on the other hand, require $X(d)$: the ethical weight of a classification error depends on whether the system allocates medical treatment or recommends films. The ethically relevant relations $R(d) \subseteq I(d) \times X(d)$---the cross-product cases where both types interact---are where the ethical knowledge gap is widest and the hardest ethical questions exist. This differentiation also clarifies where the positive side of the ethical knowledge gap is most consequential: domain-specific values are not only where harms are hardest to anticipate but also where opportunities for software to actively support autonomy, flourishing, or domain-appropriate goods are most easily missed, since realizing these goods requires precisely the integration of implementation and contextual knowledge that the gap prevents.

The consequence for institutional design is that different mechanisms dominate for different value types, and the translation layer must be calibrated accordingly. For techno-generic values, the interpretive deficit is the dominant mechanism, and it is in principle tractable: fairness concerns intrinsic to ML systems can be taught as part of engineering education and built into the sensemaking apparatus through toolkits, model cards, and bias audits. The primary intervention is sensemaking enrichment. For domain-specific values, dispersal dominates and the problem is structurally harder: the developer cannot know from the code alone that the system processes medical records or will be used for criminal identification. The primary intervention is knowledge integration---institutional mechanisms that bring contextual knowledge to developers and implementation knowledge to domain experts. For the cross-product cases, all three mechanisms are active and the full translation layer is necessary.

\subsection{Agile as Partial Translation Layer}

Agile software development methods, particularly Scrum, approximate the translation layer described above---and the framework developed in this paper can explain both where agile succeeds and where it systematically falls short. The three structural features most relevant to the ethical knowledge gap are flat hierarchies, collaborative ceremonies, and iterative incrementalism. Each partially addresses one of the gap mechanisms, but each also encounters a predicted limitation.

Agile's flat hierarchies and emphasis on team autonomy create conditions that counteract credibility attenuation. When developers are structurally empowered to question assumptions and take responsibility for their choices, they are more likely to be trusted as ethical speakers, not merely relied upon as technical producers. Peer-to-peer collaboration moves the epistemic-dependence architecture away from rigid role-based credibility assignment. However, the interpretive deficit is not a scheduling problem. Agile ceremonies---retrospectives, sprint reviews, planning meetings---provide discursive settings in which sensemaking categories can be negotiated. But the existence of such settings does not by itself dissolve the categories that make ethical significance invisible. Developers sitting in a retrospective still lack the interpretive resources to recognize that last sprint's caching decision was an ethical decision, because the sensemaking apparatus they reflect through does not include ethical categories. Giving developers time to reflect does not help if the vocabulary available for reflection does not contain the relevant concepts. What is needed is not supplementation---a checklist alongside existing sensemaking---but a restructuring of the categorical boundary between ``technical'' and ``ethical'' within the working vocabulary. That is a sensemaking intervention, not an agenda item.

A particularly promising feature is Sprint~0, where domain-specific values from $X(d)$ are surfaced before formal development cycles begin. The Product Owner plays a central role here, functioning as a de facto translation layer at the intersection of stakeholder context and technical execution. Yet the PO as translation layer faces a single-point-of-failure problem that the Hayekian analysis flags directly. The PO typically holds domain context well but lacks the implementation-level tacit knowledge that developers hold. For the cross-product cases in $R(d)$---where ethically relevant relations require both implementation details and contextual knowledge simultaneously---much of $I(d)$ never reaches the PO, and the gap persists. The developer who knows about silent data-dropping or the 72-hour caching residual will not mention it in a sprint review unless they already see it as ethically relevant---which requires precisely the interpretive categories the sensemaking deficit denies them.

Finally, agile's incrementalism can reinforce the interpretive deficit over time. Short iteration cycles reinforce present-tense technical sensemaking: each sprint enacts an environment of ``what can we ship this cycle?''---precisely the temporal frame that makes ethical significance invisible. Critics such as Olsson and V\"{a}\"{a}n\"{a}nen (2021) have argued that agile's structure discourages stepping back to assess fundamental assumptions, values, and societal impacts. Sprint~0 is a response to this, but the framework suggests it needs to be recurring rather than foundational-only. The interpretive deficit does not get solved once; it reconstitutes itself with every new sprint as the technical sensemaking apparatus reasserts its dominance. Without periodic disruption of the incremental rhythm---what we might call \emph{ethical interrupts} that force a return to foundational questions---the initial work of Sprint~0 erodes as the project progresses and new implementation decisions accumulate their own unrecognized ethical weight.

%---------------------------------------------------------------------
\section{Conclusion}
%---------------------------------------------------------------------

This paper has argued that ethical software development faces a structural epistemic problem---the \emph{ethical knowledge gap}---that is more fundamental than is generally recognized. The gap arises from the interaction of three mechanisms: the Hayekian dispersal of knowledge, the interpretive deficit produced by organizational sensemaking categories, and the credibility attenuation produced by organizational epistemic-dependence architectures. Each has been recognized in cognate literatures, but their interaction in the specific context of software development has not been analyzed.

The core contribution is the structural blockage claim: absent institutional mechanisms that simultaneously address all three components, the knowledge required for ethical decision-making is systematically unavailable at the point of decision. This transforms the ethical software development problem from a question of individual virtue (``People should care more about ethics'') into a question of institutional design (``What structures can close the ethical knowledge gap?'').

The framework provides a theoretical foundation for \emph{translation-layer} approaches---institutional mechanisms that enrich sensemaking categories, restructure epistemic-dependence architectures, and enable targeted aggregation of dispersed knowledge. It also explains why single-vector interventions consistently underperform: each addresses at most one component of a three-component problem.

The ethical knowledge gap is not a contingent organizational failure. It is a structural feature of how software is made.

%---------------------------------------------------------------------

\end{document}